\documentclass[aps,11pt,onecolumn,preprintnumbers,amsmath,amssymb]{revtex4}

\usepackage[english]{babel}
\usepackage{graphicx}
\usepackage{color}

\begin{document}

\title{Position-controlled quantum emitters with reproducible emission wavelength in hexagonal boron nitride}
\author{Clarisse Fournier$^{1}$, Alexandre Plaud$^{1,2}$, S\'ebastien Roux$^{1,2}$, Aur\'elie Pierret$^{3}$, Michael Rosticher$^{3}$, Kenji Watanabe$^{4}$, Takashi Taniguchi$^{5}$, St\'ephanie Buil$^{1}$, Xavier Qu\'elin$^{1}$, Julien Barjon$^{1}$, Jean-Pierre Hermier$^{1}$ and Aymeric Delteil$^{1,*}$ 
\\ \small \itshape $^1$ Universit\'e Paris-Saclay, UVSQ, CNRS,  GEMaC, 78000, Versailles, France. \\
$^2$ Université Paris-Saclay, ONERA, CNRS, Laboratoire d'étude des microstructures, 92322, Châtillon, France. \\
$^3$ Laboratoire de Physique de l'\'Ecole Normale Sup\'erieure, ENS, Universit\'e PSL, CNRS, Sorbonne Universit\'e, Universit\'e de Paris, 75005 Paris, France \\
$^4$ Research Center for Functional Materials, 
National Institute for Materials Science, 1-1 Namiki, Tsukuba 305-0044, Japan \\
$^5$ International Center for Materials Nanoarchitectonics, 
National Institute for Materials Science, 1-1 Namiki, Tsukuba 305-0044, Japan \\
* e-mail: aymeric.delteil@uvsq.fr}

\maketitle

\section{Abstract}
Single photon emitters (SPEs) in low-dimensional layered materials have recently gained a large interest owing to the auspicious perspectives of integration and extreme miniaturization offered by this class of materials. However, accurate control of both the spatial location and the emission wavelength of the quantum emitters is essentially lacking to date, thus hindering further technological steps towards scalable quantum photonic devices. Here, we evidence SPEs in high purity synthetic hexagonal boron nitride (hBN) that can be activated by an electron beam at chosen locations. SPE ensembles are generated with a spatial accuracy better than the cubed emission wavelength, thus opening the way to integration in optical microstructures. Stable and bright single photon emission is subsequently observed in the visible range up to room temperature upon non-resonant laser excitation. Moreover, the low-temperature emission wavelength is reproducible, with an ensemble distribution of width 3~meV, a statistical dispersion that is more than one order of magnitude lower than the narrowest wavelength spreads obtained in epitaxial hBN samples. Our findings constitute an essential step towards the realization of top-down integrated devices based on identical quantum emitters in 2D materials.

\section{Introduction}
The technological control of van der Waals materials is continually expanding, motivated by the possibility of realizing increasingly complex hetero- and nanostructures of minimal thickness. The considerable variety of impacted fields of physics~\cite{novoselov16, liu16} has been including solid-state quantum optics~\cite{aharonovich16} since the discovery of single photon emission in WSe$_2$~\cite{chakraborty15, he15, koperski15, srivastava15, tonndorf15} and hBN~\cite{tran16}. In the latter material, quantum emission is associated with point defects that were long thought to be of the intrinsic kind, although carbon impurities have been shown to play a role in the structure of at least part of the observed SPEs~\cite{mendelson20}. Their emission is found to be bright, stable~\cite{martinez16, chejanovsky16} and spectrally narrow~\cite{li17,dietrich18}, and persists up to room temperature and above~\cite{kianinia17}. They however suffer from large discrepancies between their emission wavelengths, which are typically found between 550 and 850~nm~\cite{tran16acs, castelletto20}. Epitaxial hBN grown by chemical vapour epitaxy has been recently shown to lead to a narrowing of the spectral distribution down to about 20~nm (75~meV) around a centre wavelength of 585~nm~\cite{mendelson20,stern19}. Moreover, the SPEs appear in most cases at random locations in the crystal, although often preferentially close to the flake edges~\cite{choi16}. Effort towards controlling their position has included the use of focused ion beam~\cite{ziegler19}, as well as strain through exfoliation on patterned substrates~\cite{proscia18}, but the emitters obtained with these methods exhibit large variations in their number, emission wavelength and optical properties. Moreover, the latter method results in limited possibilities of subsequent integration. In the 2D material MoS$_2$, deterministic positioning with high precision ($\sim 10$~nm) has been achieved~\cite{klein19, klein21} using He ion beam, but at the current stage the generated SPEs suffer from low count rates and large linewidths, which constitutes a major drawback for applications to photonic quantum information.

Here, we demonstrate the activation of colour centres at chosen locations using the electron beam of a commercial scanning electron microscope (SEM). Electron irradiation has already been shown to increase the formation probability of the SPEs~\cite{tran16acs, choi16, duong18}, but to date has never been the basis of a process that allows to activate SPEs at preselected locations in hBN. We show that our local irradiation process activates SPE ensembles with a submicrometric precision. The SPEs exhibit a strongly reduced ensemble linewidth with respect to prior work on 2D materials. We investigate individual quantum emitters and demonstrate advantageous photophysical properties, with in particular a high stability of both fluorescence intensity and centre wavelength fluctuations. Our work paves the way to top-down fabrication of integrated devices based on SPEs in hBN.

\section{Results}
\subsection*{Generation and characterization of SPE ensembles}

\begin{figure*}[h!] 
\centering
\includegraphics[width=\textwidth]{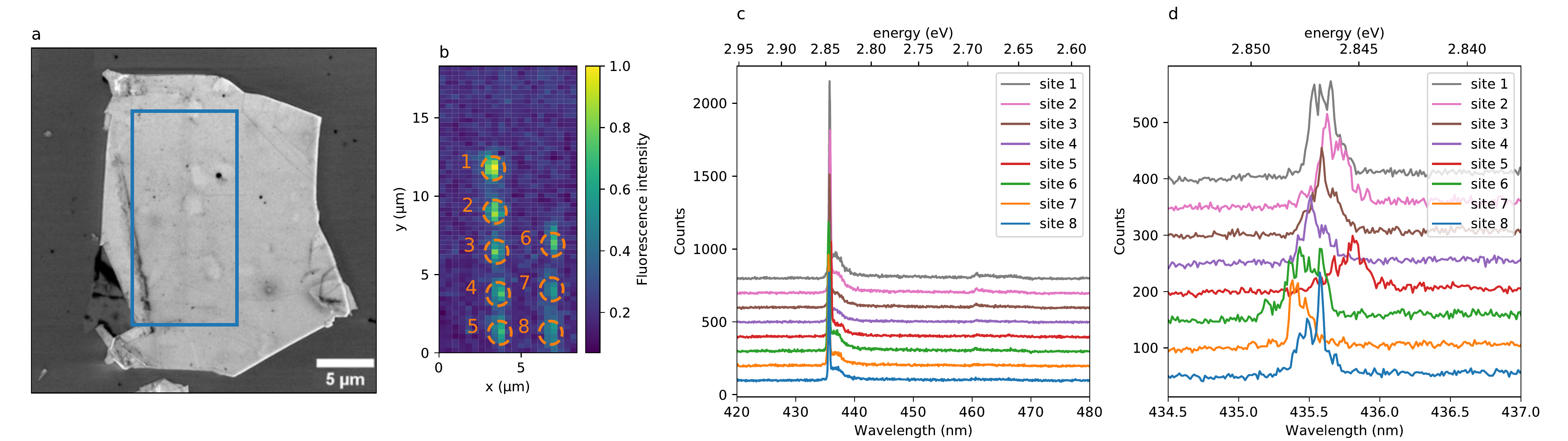}
\caption{{\bf Activation of localized ensembles of SPEs on a hBN flake.} (a) SEM image of a high-purity hBN flake of about $15 \times 20~\mu$m and 60~nm thickness. (b) Confocal map of the irradiated zone (blue rectangle in Fig.~\ref{fig01}a) with eight irradiation spots (orange dashed lines). (c) and (d) Low-temperature spectra of the eight spots with two different spectral resolutions, showing a reproducible ZPL within 0.7~nm.} \label{fig01}
\end{figure*}

We use high purity hBN synthesized at high pressure, high temperature (HPHT)~\cite{taniguchi07}, of which we exfoliate single flakes of a few tens of nanometres thickness on a silicon substrate, either with or without a top 285~nm SiO$_2$ layer. The flakes are irradiated using an electron beam of 15~keV acceleration voltage, under a current of 10~nA. We first focus on the sample with the SiO$_2$ epilayer, which we refer to as sample~1. For this sample, the beam is adjusted to be about 330~nm diameter, to compromise between maximizing both the interaction cross-section and the localization accuracy. The irradiation time is fixed at 1000~s per irradiated spot. No additional treatment is performed to the sample. After the irradiation process, the sample is subsequently characterized in photoluminescence (PL) in a confocal microscope, either at room temperature or at cryogenic temperature down to 5~K. The SPEs are non-resonantly excited using a laser at 405~nm, in pulsed or continuous wave regime. Fig.~\ref{fig01}a shows a SEM image of one of the irradiated flakes (of thickness 60~nm), together with a low temperature (5~K) confocal fluorescence map of the irradiated zone (Fig.~\ref{fig01}b). Emission from ensembles of colour centres is observed in all irradiated spots, within a radius close to that of the electron beam (see Supplementary note~1) and thus showing that the emitters are localized in a volume of about $3.5\cdot10^{-2}$~$\mu$m$^3 \approx 0.4 \lambda^3$, where $\lambda \approx 435$~nm is the emission wavelength. The low-temperature spectra associated with the irradiated sites are shown Fig.~\ref{fig01}c and d with two different resolutions. On the coarse resolution spectra (Fig.~\ref{fig01}c; see also Supplementary note~2), the overall common spectral shape of the SPEs can be observed: they exhibit a sharp zero-phonon line (ZPL) around 2.846~eV (435.7~nm) that concentrates about 40~\% of the light emission, as well as an adjacent acoustic phonon sideband (45~\%) and two phonon replica, respectively red-shifted by 155 and 185~meV (15~\%). The high resolution spectra, centred around the ZPL, is shown Fig.~\ref{fig01}d, where ensembles of discrete lines are observed. Keeping the above-mentioned irradiation parameters, we have overall realized 26~irradiation spots on 3~flakes. All of them gave rise to small ensembles of similar emission wavelength. The ensemble distribution, inferred from the PL spectra of all 26~spots, has a full width at half maximum (FWHM) of 3~meV (see Supplementary note~3), which is an order of magnitude narrower than the state of the art in 2D~materials~\cite{klein21}. We estimate the number of emitters per site to be of order of a few tens, as confirmed by photon correlation measurements (see Supplementary note~4). Remarkably, no colour centre, neither at 435~nm nor in the more usual wavelength range 550-850~nm, has been observed elsewhere on the flakes, although broad emission can be measured near the edges or close to flake defects. Interestingly, we note that light emission around 435~nm has already been observed in hBN as reported by Shevitski \textit{et al.}~\cite{shevitski19}. In the latter work however, blue emission could solely be observed in cathodoluminescence and did neither respond to laser excitation, nor exhibit any antibunching behaviour in the photon statistics. Nonetheless, it is likely that the SPEs we report here are of the same nature -- we presume that, in our case, we are able to activate the response of the emitters to photoluminescence owing to our electron irradiation parameters being very different from those used in~\cite{shevitski19}, where the electron irradiation dose is several orders of magnitude smaller. The necessity of a relatively high dose is compatible with the scenario of a dissociation of a pre-existing defect induced by the electron beam, followed by a sufficient migration of the produced species to lead to a stable optically active defect. We also mention that our irradiation procedure did not lead to SPE activation in other sources of hBN grown at atmospheric pressure (see Methods), consistently with Shevitski \textit{et al.}~\cite{shevitski19}, suggesting a physical origin of the SPEs related either to the HPHT growth conditions or to the specific solvent precursor used during the hBN synthesis.

\subsection*{Individual SPE photophysics}
\begin{figure*}[h!] 
\centering
\includegraphics[width=\textwidth]{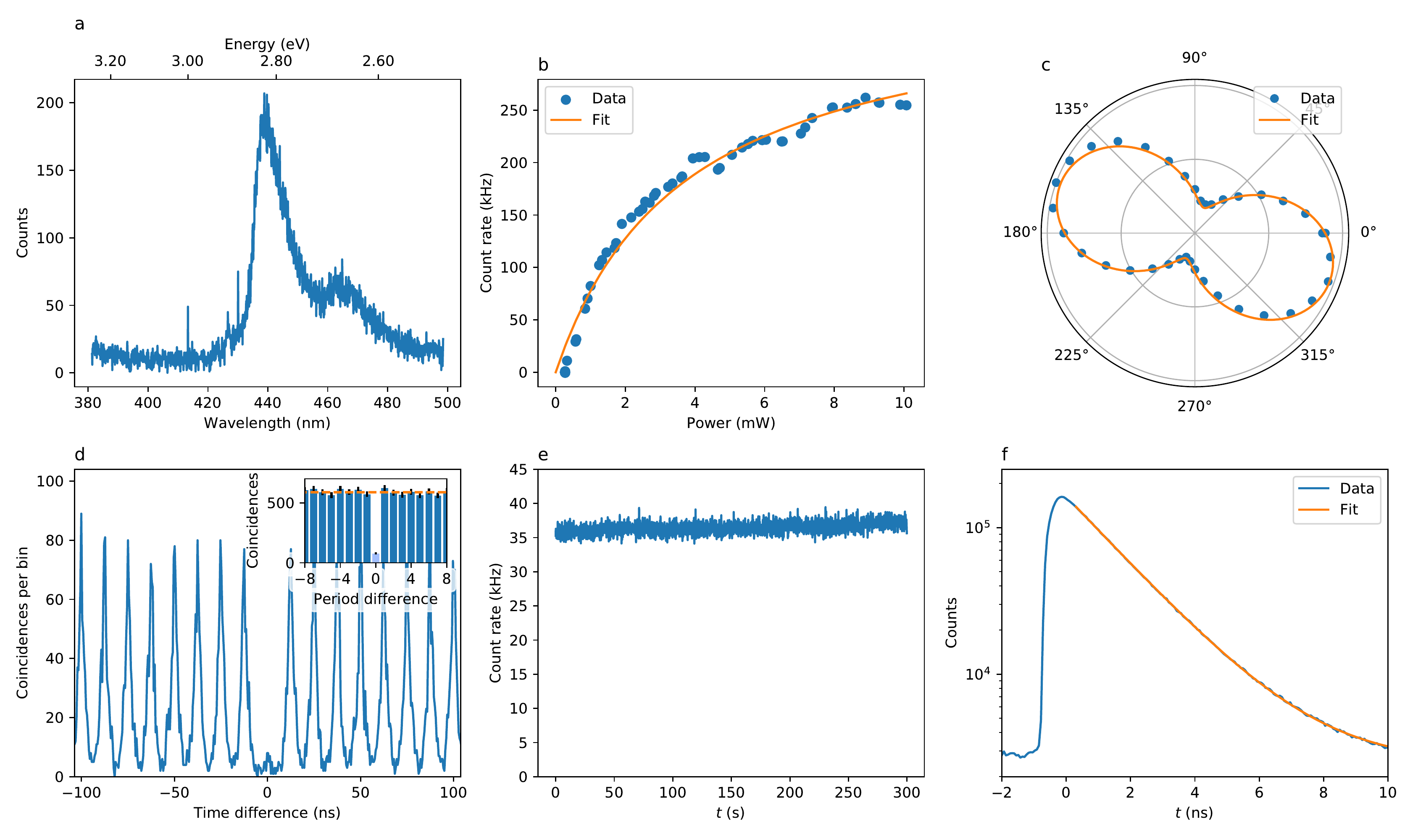}
\caption{{\bf Photophysics characterization of an individual SPE at room temperature.} (a) Emission spectrum of SPE$_1$, showing a main peak centred at 440~nm (ZPL) and a phonon replica at 465~nm. (b) Count rate as a function of the laser power in cw regime. The orange curve is a fit to the data, from which we extract a saturation power of 3.7~mW and a maximum photon detection rate of $3.6 \cdot 10^5$~Hz. (c) Count rate as a function of the angle of a polarizer placed before the detector, showing linearly polarized emission. The orange curve is a sine fit of the data. (d) Photon correlations in pulsed regime measured with 315~$\mu$W excitation power and 80~MHz repetition rate, yielding $g^{(2)}(0) = 0.12 \pm 0.01$ and thus demonstrating single photon emission. Inset: period-wise integrated coincidences (error bars: 1~standard deviation). The dashed orange line denotes the classical limit. (e) Time trace of the photon detection rate with 100~ms binning, calculated from the same raw data as (d). (f) Fluorescence decay in logarithmic scale, extracted from the same raw data as (d). The orange curve is an exponential fit to the data, yielding $\tau = 1.85$~ns.} \label{fig02}
\end{figure*}

In order to investigate the individual properties of the colour centres, we have performed additional irradiations on sample~2, with a reduced exposition time (either 300 or 600~s) and a slightly larger electron beam ($\sim 1~\mu$m diameter) on a thinner flake ($\sim 30$~nm thickness). The irradiations yielded SPEs, some of which we could characterize individually (see Supplementary note~5). The emitted light is collected by an air objective of NA 0.95, and detected using avalanche photodiodes or a spectrometer (see Methods). Fig.~\ref{fig02} shows the typical room temperature photophysical properties of a single representative colour centre, termed SPE$_1$. The emission spectrum (Fig.~\ref{fig02}a) shows that the emission mainly occurs in a ZPL centred at 440~nm, slightly red-shifted as compared with the low temperature emission. The linewidth of the ZPL is 12~nm. An optical phonon replica is visible around 465~nm. Fig.~\ref{fig02}b shows the count rate as a function of the laser power. The emission exhibits a saturation behaviour characteristic of two-level systems. We detect up to $\sim 2.5 \cdot 10 ^5$~photons per second when the SPE is excited above saturation. We fit the data with the standard power dependence of a two-level system fluorescence $I(P) = I_\mathrm{sat}/ (1 + P_\mathrm{sat}/P)$, yielding a saturation power of 3.7~mW and a saturation count rate of 0.36~MHz. This value is limited by the predominant emission of the SPE towards the high index absorptive silicon substrate and could be improved by a factor $\sim 10$ by collecting through a transparent substrate using an oil immersion objective, or by integrating the SPEs in a photonic structure. Fig.~\ref{fig02}c shows the emission polarization data of SPE$_1$. The emission is linearly polarized, suggesting a single dipole transition linearly oriented in the basal plane of the hBN crystal. We have performed second-order correlation measurements in pulsed regime to establish the quantum character of light emitted by SPE$_1$. Fig.~\ref{fig02}d shows the results we obtained, with a value of $g^{(2)}(0) = 0.12 \pm 0.01$ without background correction. This clear antibunching unequivocally demonstrates single photon emission from the colour centre. The count rate is stable over time, as can be observed on Fig.~\ref{fig02}e, with no blinking or bleaching observed at timescales $\geq 1$~ms. Absence of blinking at shorter timescales is ensured by second order correlations at intermediate timescales (see Supplementary note~7). Finally, Fig.~\ref{fig02}f shows a fluorescence decay measurement, together with an exponential fit of the data. The lifetime of the excited state is found to be $1.85$~ns, of the same order of magnitude as other families of SPEs in hBN.

\subsection*{Statistical dispersion of individual SPE properties at room temperature}
\begin{figure*}[h!] 
\centering
\includegraphics[width=0.35\textwidth]{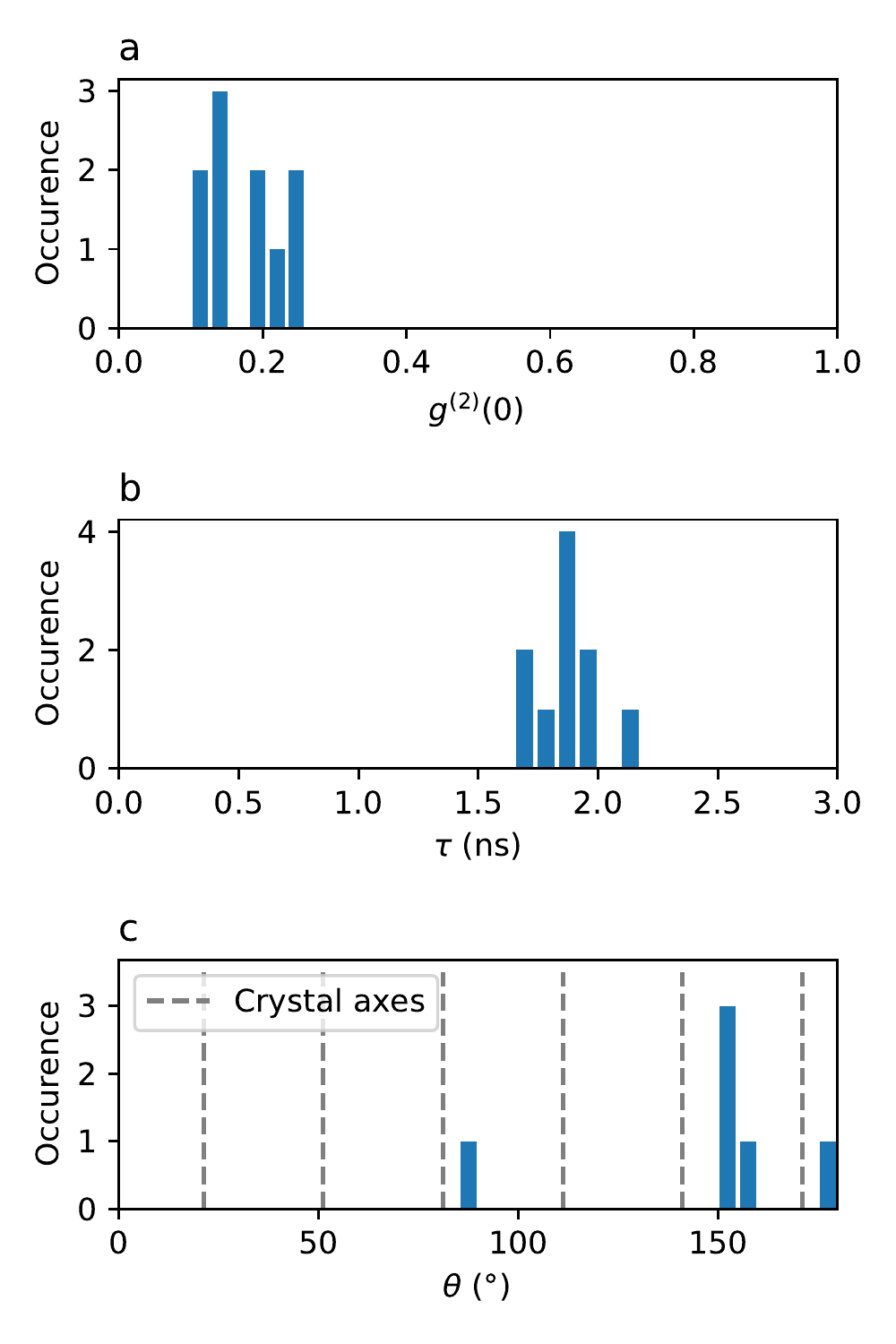}
\caption{{\bf Statistical dispersion of individual SPE properties.} (a) $g^{(2)}(0)$ of 10~SPEs, showing single-photon emission. (b) Fluorescence lifetime $\tau$ of the same 10~individual SPEs, with a mean value of 1.87~ns. (c) Polarization axis of the emission of 6~individual SPEs on the same flake, showing that most SPEs emit with a similar polarization direction.} \label{fig03}
\end{figure*}

We have performed similar measurements on 10~SPEs on two flakes (labelled SPE$_1$ to SPE$_{10}$). Fig.~\ref{fig03} shows the statistical dispersion of the associated physical quantities. The value of $g^{(2)}(0)$ (without background correction) is found between 0.1 and 0.25 as shown Fig.~\ref{fig03}a, mainly limited by fluorescence background and emission from nearby SPEs. Fig.~\ref{fig03}b shows the statistical spread of the fluorescence lifetime, which is centred around 1.87~ns with a standard deviation of 0.14~ns. Finally, the polarization angle of the emission from 6~SPEs on the same flake, to ensure a common crystalline orientation, is shown~Fig.~\ref{fig03}c. Although their directions seem correlated, they do not coincide with crystal axes. Additionally, we note that we did not observe any measurable variation of the ZPL wavelength at room temperature. These results show that the irradiation process yields SPEs with considerably homogeneous properties.

\subsection*{Low-temperature spectroscopy of individual SPEs}
\begin{figure*}[h!] 
\centering
\includegraphics[width=0.65\textwidth]{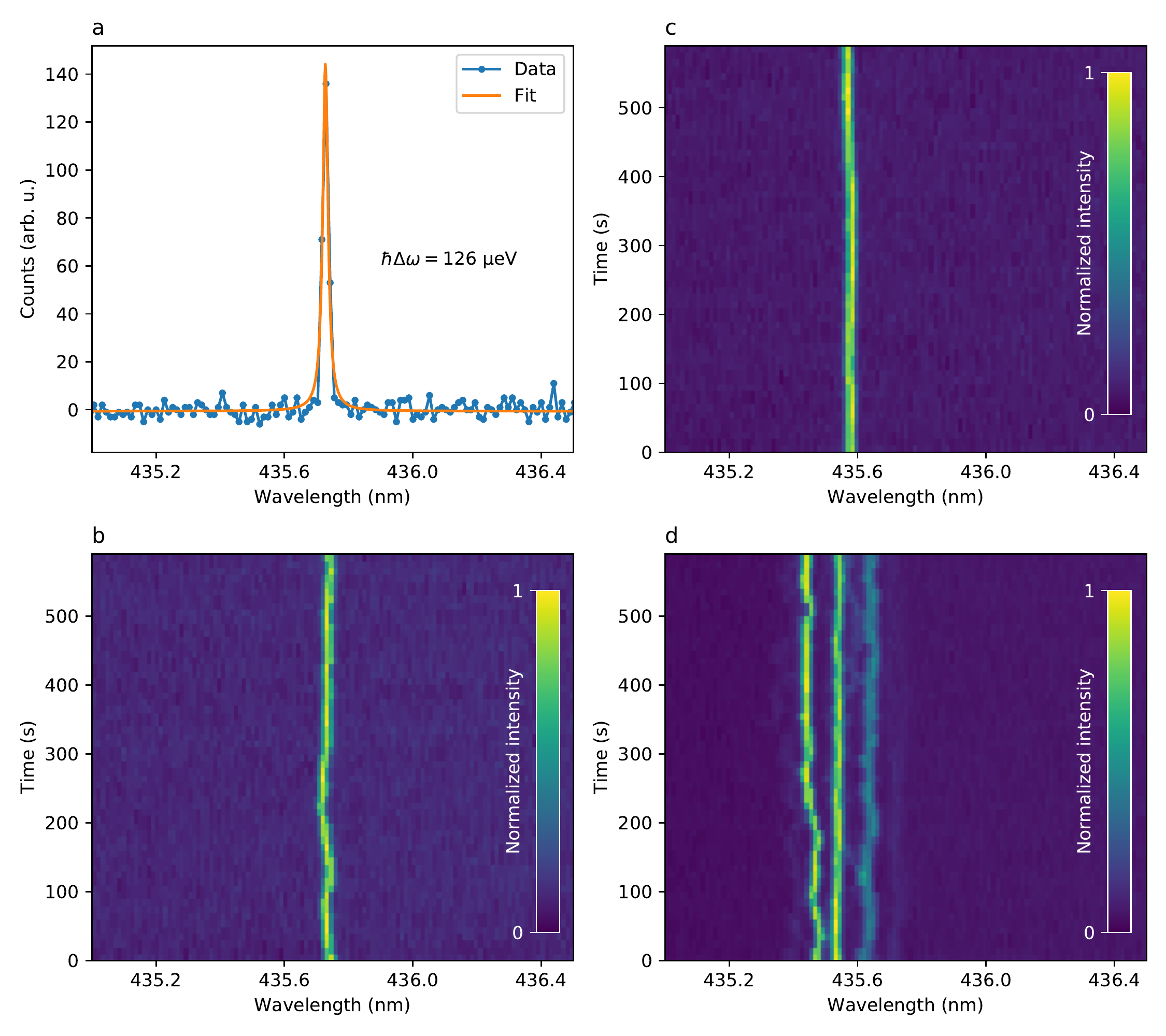}
\caption{{\bf Spectral properties of individual SPEs at low temperature (5~K).} (a) High-resolution spectrum of SPE$_1$ ZPL, showing a resolution-limited line at 435.73~nm (b) Spectral diffusion of SPE$_1$ ZPL during 600~s. The standard deviation of the centre wavelength over time is found to be 45~$\mu$eV, as determined by Lorentzian fits of the data. (c) Spectral diffusion of another SPE (SPE$_2$) and (d) spectral diffusion of an ensemble of three SPEs, with uncorrelated fluctuations of different magnitudes. All SPEs are excited with 1~mW cw laser light at 405~nm.} \label{fig04}
\end{figure*}

The spectral properties of individual SPEs at low temperature have been further investigated, and are depicted Fig.~\ref{fig04}. For most SPEs, the ZPL linewidth appears to be limited by our spectrometer resolution ($\sim 150~\mu$eV), which is the case for SPE$_1$ as shown Fig.~\ref{fig04}a. By measuring the emission spectrum as a function of time, we are able to observe the spectral diffusion of the ZPL. Fig.~\ref{fig04}b shows the result in the case of SPE$_1$. We can observe fluctuations of the centre wavelength at timescales of a few seconds, with a standard deviation of 45~$\mu$eV. The spectral diffusion of other SPEs is shown Fig.~\ref{fig04}c and d. The standard deviation of the line positions over time typically lies in the range 10 to 50~$\mu$eV (2.5 to 12~GHz), although some SPEs with larger fluctuations (a few 100s of~$\mu$eV) have also been encountered. These values are in the very low range of values usually observed for SPEs in hBN under non-resonant excitation, and could be further improved using resonant excitation~\cite{dietrich18, konthasinghe19}. The spectral diffusion, attributed to charge fluctuations in the close environment of the defect, suggests that the emission is sensitive to static electric field, thus opening the way to dc-Stark tuning of the emission line using, for instance, graphene electrodes~\cite{noh18}. Given the natural spectral proximity of the emission from different SPEs, the possibility to electrically tune the emission wavelength could potentially allow to bring any pair of SPEs to resonance, enabling quantum interference of photons emitted by distinct SPEs.

\section{Discussion}
In summary, we have demonstrated the possibility to activate SPEs in high-purity hBN at deterministic locations using the electron beam of a commercial SEM. This accessible process is well adapted to potential large-scale or industrial applications. The photophysical properties of the SPEs are advantageous and substantially replicable. In particular, the reproducibility of the emission line has no equivalent in 2D materials, and could open the way to quantum interference between distinct emitters. The relatively short emission wavelength opens the way to miniaturized on-chip applications, while still lying in the technology-friendly visible range. At low temperature, the spectral mismatch between the ZPL and the acoustic phonon sideband opens the way to the demonstration of indistinguishable photon emission by filtering out the incoherent contribution. However, while we have shown that antibunching persists up to room temperature, the emission becomes incoherent. Therefore, demonstration of room-temperature photon indistinguishability would imply to reach non-trivial cavity quantum electrodynamics regimes~\cite{Grange15, Choi19} that would entail an accurate coupling of the SPEs to a microcavity. Our work brings fundamental questions on the precise nature of the colour centres and on the physical mechanism that renders them optically active upon electron irradiation, that will motivate both further experimental investigations and theoretical studies. On the technological side, it will be desirable to settle methods allowing to deterministically obtain a single SPE per irradiation spot. Such process could for instance make use of in-situ cathodoluminescence measurements~\cite{schue19} during the irradiation process, heralding successful activation of a colour centre. This could in turn enable deterministic coupling of individual SPEs to photonic~\cite{kim18} or plasmonic~\cite{tran17} nanostructures. We expect our research to bring new possibilities to the field of quantum optics in 2D materials, that could yield applications in nanophotonics, integrated quantum optics and quantum information science.

\section*{METHODS}

\textbf{Sample fabrication.} 
High-purity hBN was grown under high pressure/high temperature using barium boron nitride (Ba$_3$B$_2$N$_4$) as a solvent system.
The hBN flakes were obtained by mechanical exfoliation of bulk material on commercial silicon substrates. We used two different exfoliation methods for the two samples: for sample~1, the hBN has been exfoliated using two 3~mm thick polydimethylsiloxane (PDMS) stamps on a SiO$_2$/Si substrate (with 285~nm SiO$_2$ epilayer), and for sample~2 we used Scotch tape to exfoliate on a Si substrate. This allows to rule out the role of a specific residue in the SPE creation process. Prior to the exfoliation, the substrates were cleaned using acetone for 5~minutes, isopropyl alcohol for 5~minutes, followed by 5~minutes of 30~W oxygen plasma treatment. Two control samples grown at atmospheric pressure have also been used: a APHT (atmospheric pressure, high temperature) sample grown in KSU (Kansas, USA) using Ni/Cr solvent~\cite{kubota07}, and a sample grown in LMI (Lyon, France) using PDC (polymer derived ceramics)~\cite{matsoso20}. Both samples have been exfoliated using adhesive tape on a SiO$_2$/Si substrate. The SEM imaging and the electron irradiations were performed in a commercial SEM (JEOL 7001F). The flake thicknesses were measured with an atomic force microscope.

\textbf{Optical characterization.} 
For room temperature characterization, the sample was placed in a confocal microscope with an air objective of NA 0.95. Low-temperature characterization was done in a closed-cycle cryostat and a low-T objective of NA 0.8 was used. In both cases, the sample was placed on three-axis piezo positioners. A 405~nm laser diode was used to excite the SPEs, either in continuous wave or in pulsed regime (pulse length $\sim 200$~ps, repetition rate 80~MHz). A dichroic mirror (cutoff wavelength 414~nm) and a fluorescence filter allowed to suppress back-reflected laser light. The signal was fibre-coupled to either a grating spectrometer (Princeton Instruments) or avalanche photodiodes (Micro Photon Devices) with 30~\% collection efficiency in the relevant wavelength range, and the detection event were recorded using a time-tagged single photon counting module (PicoQuant). In the photon correlations measurements, only the photons emitted after the laser pulse have been recorded in order to avoid double excitation events caused by the finite laser pulselength. 

{}

\vspace{1 cm}

\textbf{Supplementary Information} can be found at the end of the present document. \\ 

\textbf{Acknowledgements} The Authors acknowledge many useful discussions with Christophe Arnold. The authors thank Bruno Berini for atomic force microscope measurements and Christ\`ele Vilar for technical support on electron microscopy. This work is supported by funding from the French Institute of Physics (INP), and from the French national research agency (ANR) under grant agreement No ANR-14-CE08-0018 (GoBN: Graphene on Boron Nitride Technology). This work also received funding from the European Union's Horizon 2020 research and innovation program under Grant Nos. 785219 (Graphene Flagship Core~2) and 881603 (Graphene Flagship Core~3). K.W. and T.T. acknowledge support from the Elemental Strategy Initiative conducted by the MEXT, Japan, Grant Number JPMXP0112101001, JSPS KAKENHI Grant Number JP20H00354 and the CREST (JPMJCR15F3), JST.\\

\textbf{Author Contributions} K.W. and T.T. grew the hBN.  Al.P. and J.B. discovered the SPEs in cathodoluminescence. Au.P. and M.R. fabricated the samples. Al.P., S.R. and J.B. designed the irradiation protocol and performed the irradiations. C.F. and A.D. performed the optical measurements. S.B., X.Q., J.P.H. and A.D. designed the optical experiments and discussed the data. A.D. supervised the project and wrote the paper, with input from all authors.\\

\textbf{Competing Interests} The authors declare no competing interests.\\

\textbf{Data availability} The data generated in this study are available at https://doi.org/10.5281/zenodo.4768457.\\

\newpage

\title{Supplementary Information \\~\\ Position-controlled quantum emitters with reproducible emission wavelength in hexagonal boron nitride}

\author{Clarisse Fournier, Alexandre Plaud, S\'ebastien Roux, Aur\'elie Pierret, \\
 Michael Rosticher, Kenji Watanabe, Takashi Taniguchi, St\'ephanie Buil, \\
  Xavier Qu\'elin, Julien Barjon, Jean-Pierre Hermier and Aymeric Delteil}


\baselineskip20pt


\maketitle

\addto\captionsenglish{\renewcommand{\figurename}{Supplementary figure}}
\setcounter{figure}{0}

\begin{center}
\huge \textbf{Supplementary Information} \\ ~ \\
\large
\textbf{Position-controlled quantum emitters with reproducible emission wavelength in hexagonal boron nitride} \\ ~ \\
\normalsize
Clarisse Fournier, Alexandre Plaud, S\'ebastien Roux, Aur\'elie Pierret, \\
 Michael Rosticher, Kenji Watanabe, Takashi Taniguchi, St\'ephanie Buil, \\
  Xavier Qu\'elin, Julien Barjon, Jean-Pierre Hermier and Aymeric Delteil
  
\end{center}
\section*{Supplementary note 1: Characterization of the spot diameter}

\begin{figure*}[h!] 
\centering
\includegraphics[width=0.8\textwidth]{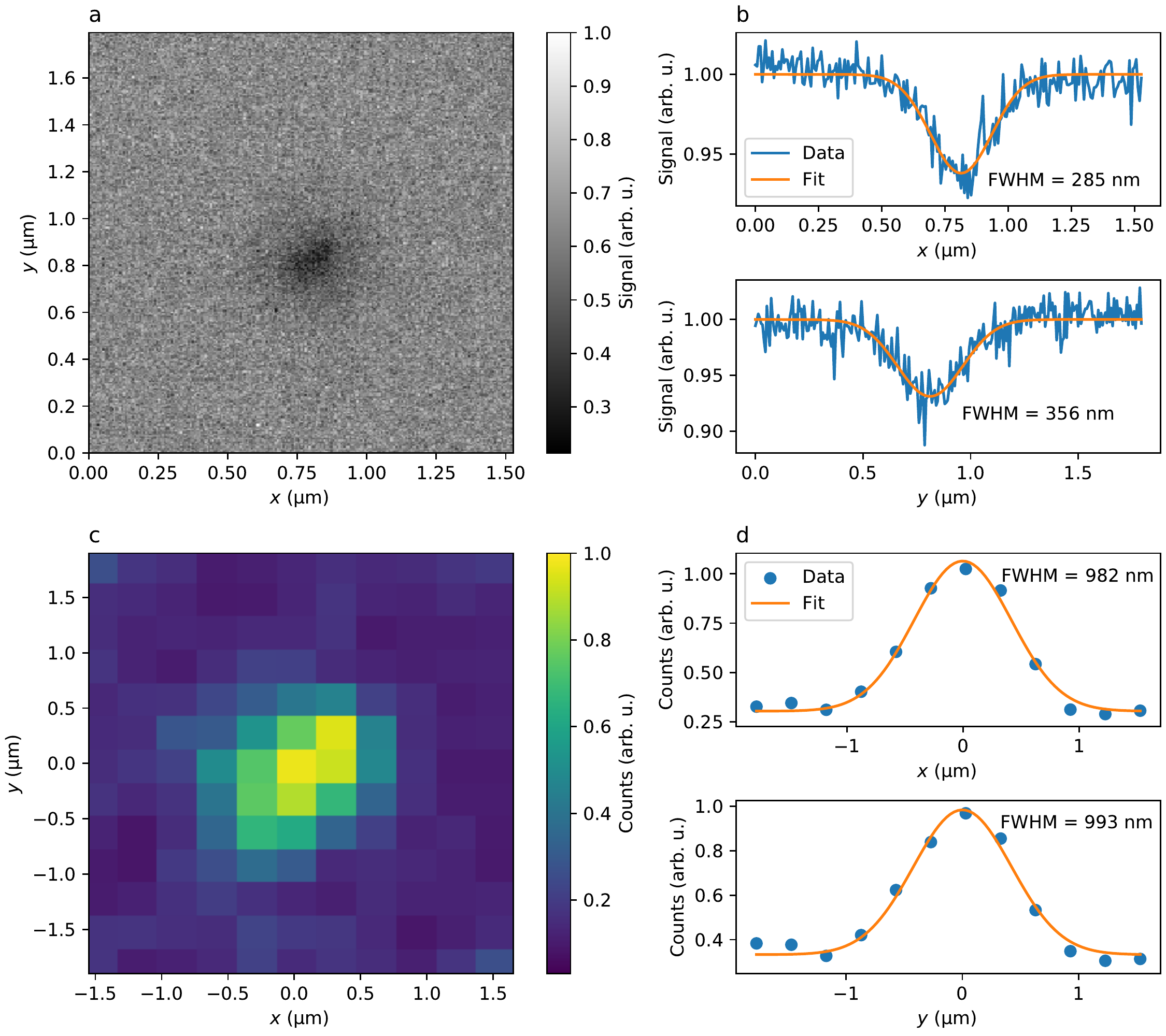}
\caption{(a) SEM image of an irradiation on the SiO$_2$ substrate. (b) Sum along $x$ (upper panel) and $y$ (lower panel) from which we estimate the beam dimensions. The orange lines are Gaussian fits of the data. (c) Room temperature photoluminescence confocal map of a SPE ensemble from an irradiation spot. (d) Cuts along $x$ (upper pannel) and $y$ (lower pannel) from which we extract the characteristic dimensions of the SPE ensemble.} \label{figS1}
\end{figure*}

In the first round of irradiation (sample~1, corresponding to the figure~1 of the main text), the electron beam diameter has been calibrated by irradiating the substrate and imaging the modification it yielded to the secondary electron imaging signal in the SEM (supplementary figure~\ref{figS1}a). The irradiation leads to a reduction of the signal from the substrate within a slightly elliptic Gaussian spot. We summed the spot image along two orthogonal direction and fitted the result to obtain a mean FWHM of 315~nm (supplementary figure~\ref{figS1}b).

The spatial distribution of the PL signal from SPE ensembles has been characterized with a similar procedure: the room-temperature PL signal (supplementary figure~\ref{figS1}c) was summed along two orthogonal direction and the resulting Gaussian shape was fitted, yielding the FWHM of the intensity distribution. Statistics from 10~irradiation spots lead to an average FWHM of 850~nm. This value is slightly greater than the spot size convolved by the emission wavelength (which yields 550~nm). The mismatch could be attributed to the relatively long duration of the irradiations leading to drifts in the beam position and therefore to an effective broadening of the spot, as well as to secondary electrons creating colour centres in the close vicinity of the spot.

\clearpage

\section*{Supplementary note 2: Spectral shape and phonon replica}

\begin{figure*}[h!] 
\centering
\includegraphics[width=0.75\textwidth]{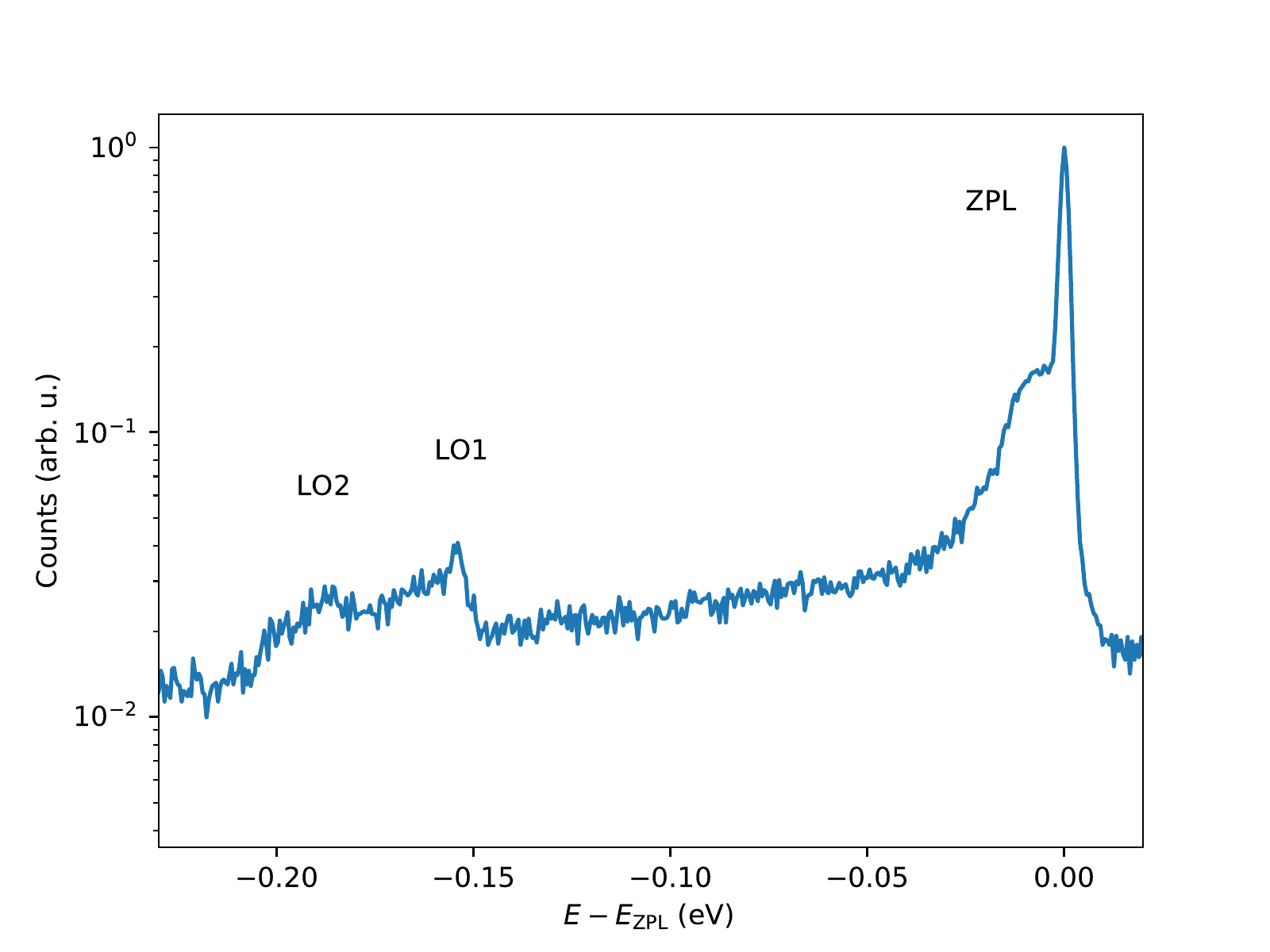}
\caption{Ensemble spectrum plotted as a function of the relative energy from the ZPL maximum ($E_\mathrm{ZPL}$).} \label{figS2}
\end{figure*}

In order to better appreciate the spectral lineshape of the SPEs, we show supplementary figure~\ref{figS2} an ensemble spectrum -- calculated as the sum of the spectra shown figure~1c of the main text -- in logarithmic scale. We can distinguish an acoustic phonon sideband~\cite{vuong16}, as well as two phonon replica, labelled LO1 and LO2, respectively redshifted by 155~meV and 185~meV relatively to the ZPL. We note that these values are slightly lower than usually observed for other SPEs~\cite{wigger19}, with however a similar splitting of about 30~meV between LO1 and~LO2.

\clearpage

\section*{Supplementary note 3: Ensemble linewidth}

\begin{figure*}[h!] 
\centering
\includegraphics[width=0.6\textwidth]{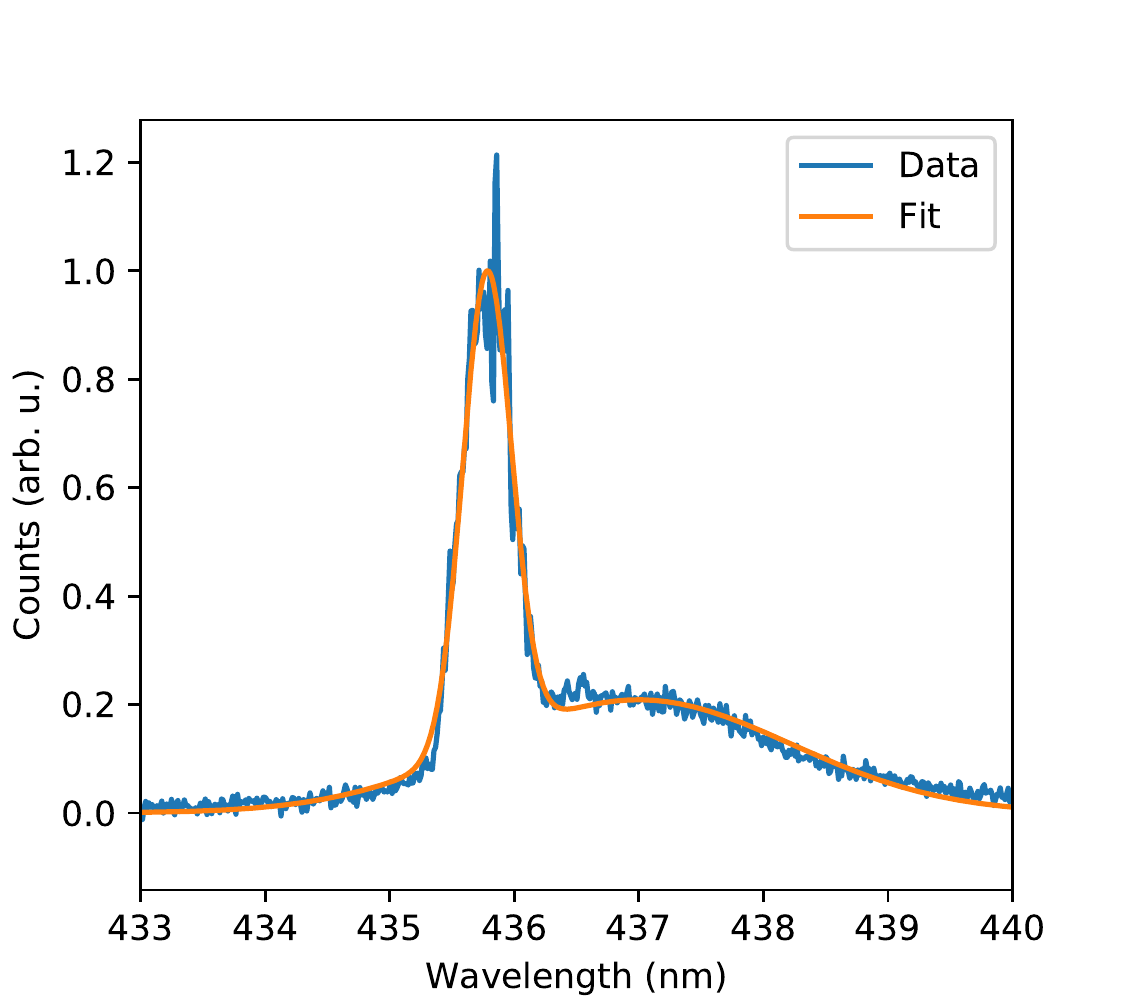}
\caption{Ensemble spectrum summed over 26~irradiation sites on sample~1. The Gaussian fit provides a 3~meV FWHM for the ZPL distribution.} \label{figS3}
\end{figure*}

To estimate the probability distribution of the ZPL wavelength, we have summed the spectra taken over the 26~irradiations we performed on sample~1. This corresponds to several hundreds of emitters (see next section). The ensemble spectrum we obtain (supplementary figure~\ref{figS3}) was fitted with a sum of two Gaussian functions, respectively associated with the ZPL and the acoustic phonon sideband. The fit parameters provide the ZPL centre wavelength (435.78~nm) and FWHM (0.46~nm, \textit{i.e.} 3.0~meV). The phonon sideband has a centre wavelength of 437.0~nm and a FWHM of 2.9~nm, \textit{i.e.} 19~meV.

\clearpage

\section*{Supplementary note 4: Estimation of SPE number in sample~1}

\begin{figure*}[h!] 
\centering
\includegraphics[width=\textwidth]{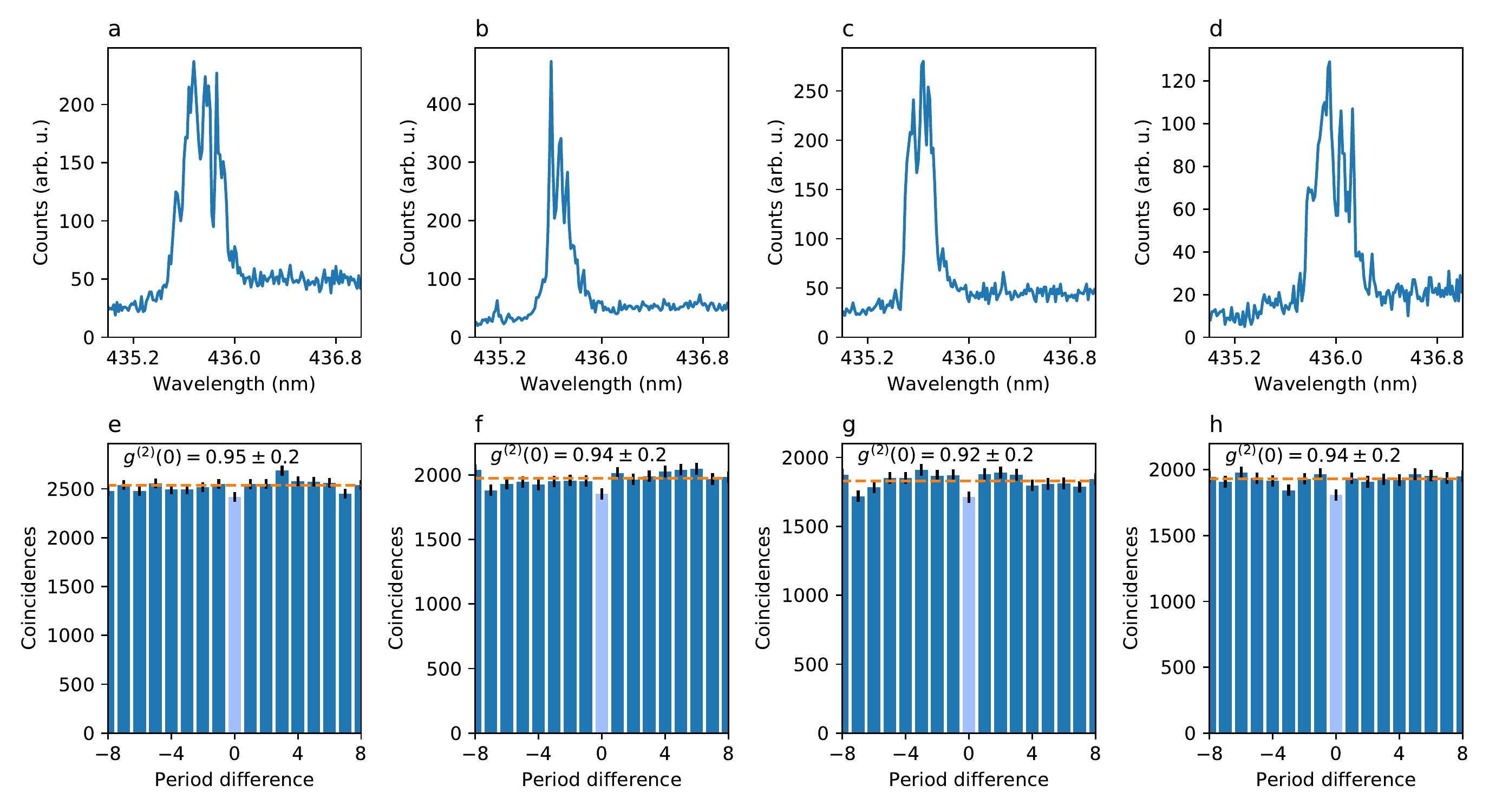}
\caption{(a)-(d) Low-temperature PL spectra of four SPE ensembles and (e)-(f) corresponding RT second-order correlation demonstrating some degree of antibunching. The orange dashed line denotes the classical limits. The value of $g^{(2)}(0)$ is indicated above each histogram. The error bars represent 1~standard deviation as calculated from Poissonian statistics of the photon counting events.}. \label{figS4}
\end{figure*}

The number of emitters in the irradiated spots can be estimated by combined low-temperature PL spectroscopy and second-order correlation of the ensemble emission. Fig~\ref{figS4}a-d shows PL spectra of four typical ensembles at 5~K. Although counting the number of SPEs is not directly possible due to the too close proximity of the ZPLs, several discrete lines can be observed, suggesting that the number of individual lines is at least of a few units. Fig~\ref{figS4}e-h shows the room-temperature second order correlation of the same ensembles, with $g^{(2)}(0)$ typically found around 0.95. As a reference, a $g^{(2)}(0) = 0.95 \pm 0.02$ corresponds to a number of $\sim 20$ ideal single-photon emitters -- in general, the $g^{(2)}(0)$ of $N$ ideal single-photon sources is given by $1 - 1/N$. We note that the number estimated from $g^{(2)}(0)$ constitutes an upper bound of the SPE number, since background, dark counts and other nonidealities can only degrade the antibunching.

\clearpage

\section*{Supplementary note 5: Isolation of individual SPEs}

\begin{figure*}[h!] 
\centering
\includegraphics[width=0.8\textwidth]{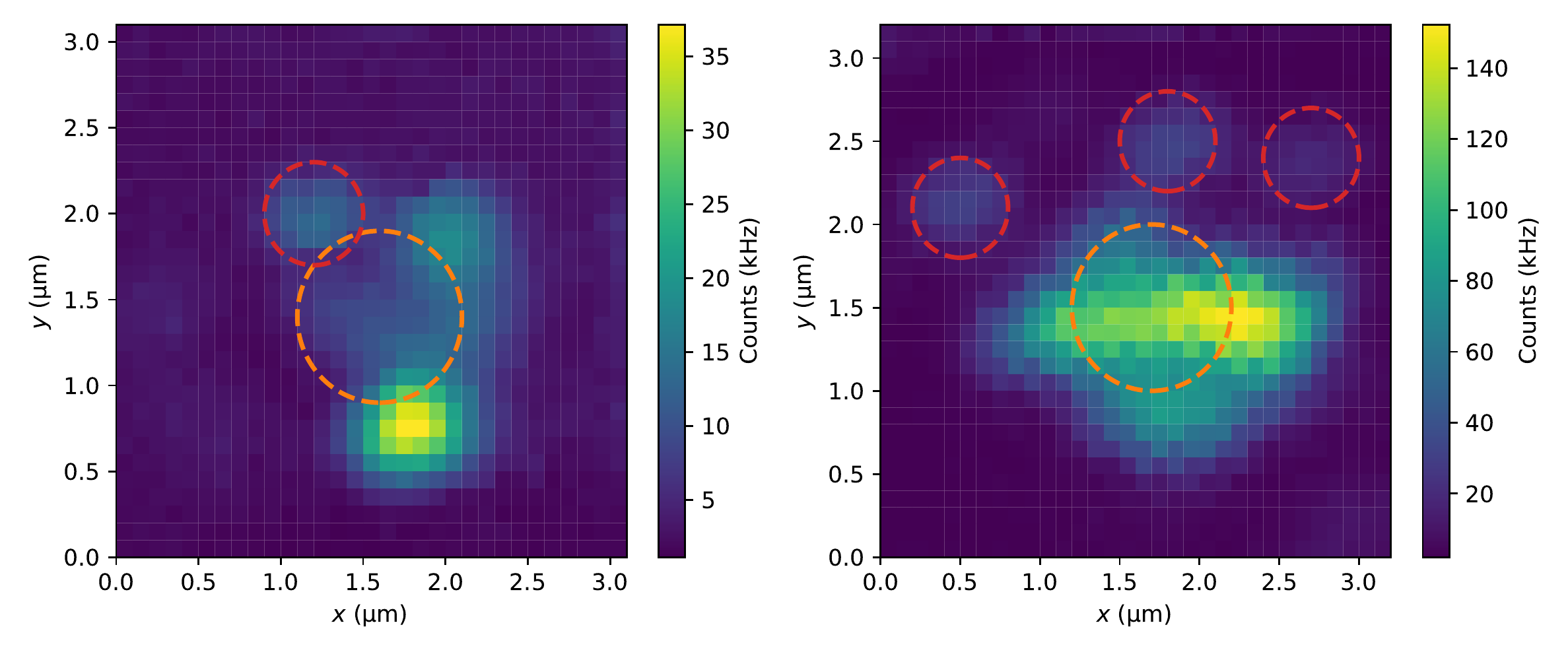}
\caption{Room-temperature confocal maps of two irradiations. The orange circle is the estimated position of the electron beam and the red circles point individual SPEs.} \label{figS5}
\end{figure*}

As mentioned in the main text, a 30~nm flake has been irradiated with a lower dose, spread over larger areas. Supplementary figure~\ref{figS5} shows a confocal map of two irradiated spots with 10~min irradiation time. The orange circles denote the nominal beam diameter (FWHM $\sim 1~\mu$m) used for this flake. With these irradiation parameters, individual SPEs can be isolated at the vicinity of the spots as indicated by the red circles. The rest of the luminescence originates from small ensembles of two SPEs or more, as confirmed by both $g^{(2)}$ measurements and spectroscopy, with the presence of several ZPLs.

\clearpage

\section*{Supplementary note 6: Flake characterization}

\begin{figure*}[h!] 
\centering
\includegraphics[width=0.75\textwidth]{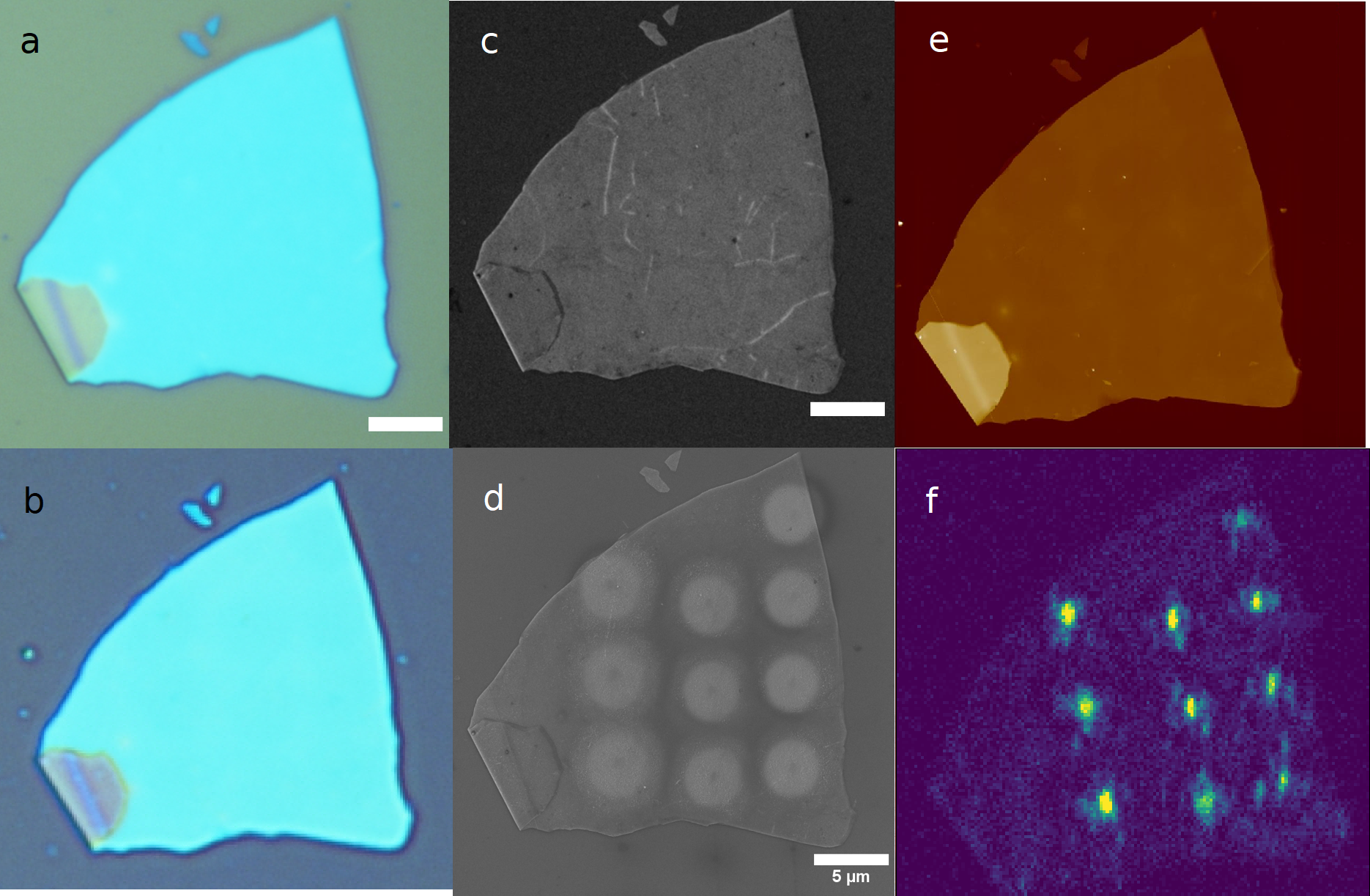}
\caption{(a) Optical microscope image of the flake before irradiation, and (b) after irradiation. (c) Scanning electron microscope before and (d) after irradiation. (e) AFM image of the flake after irradiation. (f) Low-T confocal PL scan of the flake.} \label{figS6}
\end{figure*}

On supplementary figure~\ref{figS6} we show optical microscopy, scanning electron microscopy (SEM) and atomic force microscopy (AFM) of an irradiated flake from sample~1, with 10 irradiated spots (with irradiation parameters as described in the main text). The effect of the electron irradiation is not measurable in optical microscopy. On the AFM map, the only trace of the irradiation process on the 60~nm thick flake is a slightly thinner ($\sim$~2--3~nm) zone within a 2.5~$\mu$m radius around the irradiations -- which we also observe if we directly irradiate the substrate. The SEM image, in turn, reveals a brighter contrast a couple of micrometres around the irradiation spots. The confocal PL scan reveals ten bright spots at the position of the irradiated sites.

\clearpage

\section*{Supplementary note 7: Continuous wave second order correlations}

\begin{figure*}[h!] 
\centering
\includegraphics[width=\textwidth]{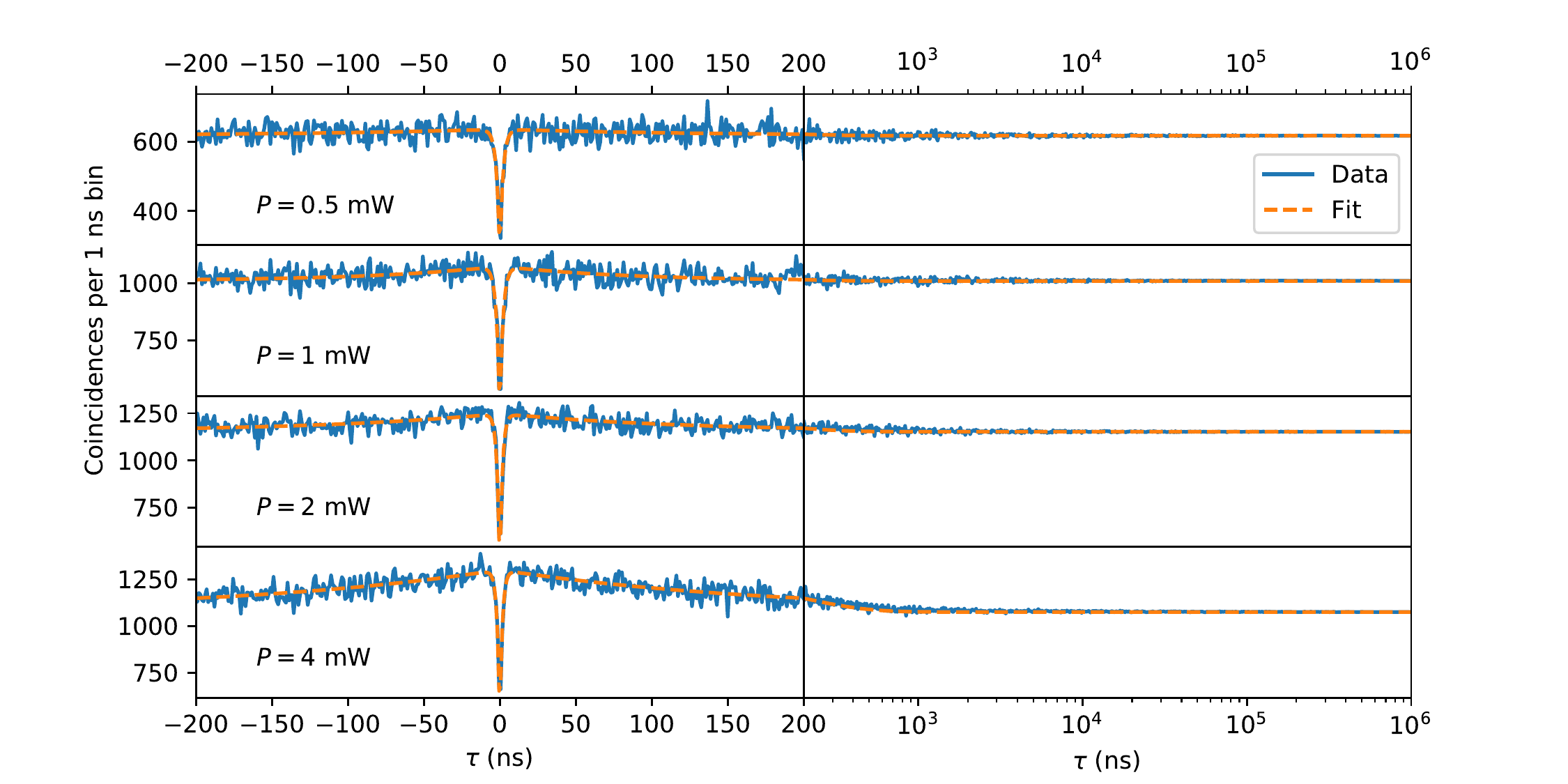}
\caption{$g^{(2)}(\tau)$ measured at four different powers. Left panel: linear scale around 0. Right panel: logarithmic scale for the time axis. The orange line is a fit with a function of the form $g^{(2)}(\tau) = A - B e^{-|\tau|/t_1} + C e^{-|\tau|/t_2}$} \label{figS7}
\end{figure*}

We have measured the second-order photon correlation of a SPE under continuous wave excitation at several powers around saturation (supplementary figure~\ref{figS7}). The short-time correlation function exhibits a power-dependent bunching contribution decaying at timescales of order 100~ns. This observation is similar to what has been observed in SPEs from other physical systems~\cite{kitson98, wang18} and is consistent with the presence of a third metastable state of lifetime $\sim 400$~ns~\cite{martinez16}. The long-time $g^{(2)}(\tau)$ is constant up to timescales of milliseconds, confirming the absence of blinking dynamics in the range ns to ms. The stability at longer timescales is established by the intensity timetraces for $P < P_{sat}$ (see figure~2 of the main text), although at the highest powers, sizeable intensity fluctuations can be observed.

{}


\section*{References}

\begin{thebibliography}{}

%
%
%
%
%
%
%
%
%



\bibitem{novoselov16} Novoselov, K. S., Mishchenko, A., Carvalho, A. \& Neto, C. 2D materials and van der Waals heterostructures. \textit{Science} \textbf{353}, aac9439 (2016). 


\bibitem{liu16} Liu, Y., Weiss, N. O., Duan, X., Cheng, H.-C., Huang, Y. \& Duan, X. Van der Waals heterostructures and devices. \textit{Nature Rev. Mater.} \textbf{1}, 16042 (2016). 



\bibitem{aharonovich16} Aharonovich, I., Englund, D. \& Toth, M. Solid-state single-photon emitters. \textit{Nature Photon} \textbf{10}, 631–641 (2016). 



\bibitem{chakraborty15}
Chakraborty, C., Kinnischtzke, L., Goodfellow, K. M., Beams, R. \&
Vamivakas, A. N. Voltage-controlled quantum light from an atomically thin
semiconductor. \textit{Nature Nanotechnol.} \textbf{10}, 507–511 (2015).

\bibitem{he15}
He, Y.-M. \textit{et al}. Single quantum emitters in monolayer semiconductors. \textit{Nature Nanotechnol.} \textbf{10}, 497–502 (2015).

\bibitem{koperski15} Koperski, M. \textit{et al}. Single photon emitters in exfoliated WSe$_2$ structures. \textit{Nature Nanotechnol.} \textbf{10}, 503–506 (2015).

\bibitem{srivastava15} Srivastava, A. \textit{et al}. Optically active quantum dots in monolayer WSe$_2$. \textit{Nature Nanotechnol.} \textbf{10}, 491–496 (2015).

\bibitem{tonndorf15} Tonndorf, P. \textit{et al}. Single-photon emission from localized excitons in an atomically thin semiconductor. \textit{Optica} \textbf{2}, 347-352 (2015).

\bibitem{tran16} Tran, T. T., Bray, K., Ford, M. J., Toth, M. \& Aharonovich, I. Quantum emission from hexagonal boron nitride monolayers. \textit{Nat. Nanotechnol.} \textbf{11}, 37–41 (2016).

\bibitem{mendelson20} Mendelson, N. \textit{et al}. Identifying carbon as the source of visible single-photon emission from hexagonal boron nitride. \textit{Nat. Mater.} (2020). 

%
%
%
%

\bibitem{martinez16} Mart\'inez, L. J., Pelini, T., Waselowski, V., Maze, J. R., Gil, B., Cassabois, G.\& Jacques, V. Efficient single photon emission from a high-purity hexagonal boron nitride crystal. \textit{Phys. Rev. B} \textbf{94}, 121405(R) (2016).

\bibitem{chejanovsky16} Chejanovsky, N. \textit{et al.} Structural Attributes and Photodynamics of Visible Spectrum Quantum Emitters in Hexagonal Boron Nitride. \textit{Nano Lett}. \textbf{16}, 11, 7037–7045 (2016). 

\bibitem{li17} Li, X. \textit{et al}. Nonmagnetic Quantum Emitters in Boron Nitride with Ultranarrow and Sideband-Free Emission Spectra. \textit{ACS Nano}, \textbf{11}, 7, 6652–6660 (2017).

\bibitem{dietrich18} Dietrich, A. \textit{et al}. Observation of Fourier transform limited lines in hexagonal boron nitride. \textit{Phys. rev. B} \textbf{98}, 081414(R) (2018).


\bibitem{kianinia17} Kianinia M., Regan B., Tawfik, S. A. \textit{et al}. Robust solid-state quantum system operating at 800~K. \textit{ACS Photonics} \textbf{4}, 768–73 (2017).

%
\bibitem{tran16acs} Tran, T. T. \textit{et al.} Robust Multicolor Single Photon Emission from Point Defects in Hexagonal Boron Nitride. \textit{ACS Nano} \textbf{10}, 8, 7331–7338 (2016). 

\bibitem{castelletto20} Castelletto, S., Inam, F. A., Sato, S. \& Boretti, A. Hexagonal boron nitride: a review of the emerging material platform for single-photon sources and the spin–photon interface. \textit{Beilstein J. Nanotechnol}. \textbf{11}, 740–769 (2020).

\bibitem{stern19} Stern, H.L. \textit{et al}. Spectrally Resolved Photodynamics of Individual Emitters in Large-Area Monolayers of Hexagonal Boron Nitride. \textit{ACS Nano} \textbf{13}, 4538-4547 (2019).

\bibitem{choi16} Choi, S. \textit{et al.} Engineering and Localization of Quantum Emitters in Large Hexagonal Boron Nitride Layers. \textit{ACS Appl. Mater. Interfaces} \textbf{8}, 43, 29642–29648 (2016).

\bibitem{ziegler19} Ziegler, J. \textit{et al.} Deterministic Quantum Emitter Formation in Hexagonal Boron Nitride via Controlled Edge Creation. \textit{Nano Lett.} \textbf{19}, 2121-2127 (2019). 

\bibitem{proscia18} Proscia, N. V. \textit{et al.} Near-deterministic activation of room-temperature quantum emitters in hexagonal boron nitride. \textit{Optica} \textbf{5}, 9, 1128-1134 (2018). 



\bibitem{klein19} Klein, J. \textit{et al.} Site-selectively generated photon emitters in monolayer MoS$_2$ via local helium ion irradiation. \textit{Nature Commun.} \textbf{10}, 2755 (2019).


\bibitem{klein21} Klein, J. \textit{et al.} Engineering the Luminescence and Generation of Individual Defect Emitters in Atomically Thin MoS$_2$. \textit{ACS Photonics} \textbf{8}, 669-677 (2021).


\bibitem{duong18} Duong, H. N. M. \textit{et al.} Effects of high energy electron irradiation on quantum emitters in hexagonal boron nitride. \textit{ACS Appl. Mater. Interfaces} \textbf{10}, 29, 24886–24891 (2018). 



\bibitem{taniguchi07} Taniguchi, T. \& Watanabe, K. Synthesis of high-purity boron nitride single crystals under high pressure by using Ba–BN solvent. \textit{J. Cryst. Growth} \textbf{303}, 525–529 (2007).


\bibitem{shevitski19} Shevitski, B. \textit{et al}. Blue-light-emitting color centers in high-quality hexagonal boron nitride. \textit{Phys. Rev. B} \textbf{100}, 155419 (2019). 

\bibitem{konthasinghe19} Konthasinghe, K. \textit{et al.} Rabi oscillations and resonance fluorescence from a single hexagonal boron nitride quantum emitter. \textit{Optica} \textbf{6}, 542-548 (2019).


\bibitem{noh18} Noh, G. \textit{et al}. Stark Tuning of Single-Photon Emitters in Hexagonal Boron Nitride. \textit{Nano Lett}. \textbf{18}, 4710-4715 (2018).


\bibitem{Grange15} Grange T., Hornecker, G., Hunger, D., Poizat, J.-P., G\'erard, J.-M., Senellart, P. \& Auff\`eves, A. Cavity-Funneled Generation of Indistinguishable Single Photons from Strongly Dissipative Quantum Emitters.
\textit{Phys. Rev. Lett.} \textbf{114}, 193601 (2015).

\bibitem{Choi19} Choi, H., Zhu, D., Yoon, Y. \& Englund, D. Cascaded Cavities Boost the Indistinguishability of Imperfect Quantum Emitters. \textit{Phys. Rev. Lett.}\textbf{122}, 183602 (2019).


\bibitem{schue19} Schu\'e, L. \textit{et al.} Bright Luminescence from Indirect and Strongly Bound Excitons in h-BN. \textit{Phys. Rev. Lett.} \textbf{122}, 067401 (2019).

\bibitem{kim18} Kim, S., Fr\"och, J.E., Christian, J. \textit{et al}. Photonic crystal cavities from hexagonal boron nitride. \textit{Nat. Commun.} \textbf{9}, 2623 (2018). 

\bibitem{tran17} Tran, T. T., Wang, D., Xu, Z.-Q., Yang, A., Toth, M., Odom, T. W., \& Aharonovich, I. Deterministic Coupling of Quantum Emitters in 2D Materials to Plasmonic Nanocavity Arrays.\textit{ Nano Lett.} \textbf{17}(4), 2634–2639 (2017). 

\bibitem{kubota07} Kubota, Y., Watanabe, K., Tsuda, O., \&, Taniguchi, T. Deep Ultraviolet Light-Emitting Hexagonal Boron Nitride Synthesized at Atmospheric Pressure. \textit{Science} \textbf{317}, 932 (2007).

\bibitem{matsoso20} Matsoso, B. \textit{et al.} Synthesis of hexagonal boron nitride 2D layers using polymer derived ceramics route and
derivatives. \textit{J. Phys. Mater.} \textbf{3}, 034002 (2020).

\end{thebibliography}

\clearpage

\begin{thebibliography}{}

\bibitem{vuong16} Vuong, T. \textit{et al.} Phonon-Photon Mapping in a Color Center in Hexagonal Boron Nitride. \textit{Phys. Rev. Lett.} \textbf{117}, 097402 (2016) .

\bibitem{wigger19} Wigger, D. \textit{et al.} Phonon-assisted emission and absorption of individual color centers in hexagonal boron nitride. \textit{2D Mater.} \textbf{6}, 035006 (2019).
 
\bibitem{kitson98} Kitson, S.C., Jonsson, P., Rarity, J.G., Tapster, P.R. Intensity fluctuation spectroscopy of small numbers of dye molecules in a microcavity. \textit{Phys. Rev. A} \textbf{58}, 620 (1998).

\bibitem{wang18} Wang, J. \textit{et al.} Bright room temperature single photon source at telecom range in cubic silicon carbide. \textit{Nature Commun.} \textbf{9}, 4106 (2018).

\bibitem{martinez16} Mart\'inez, L. J., Pelini, T., Waselowski, V., Maze, J. R., Gil, B., Cassabois, G. \& Jacques, V. Efficient single photon emission from a high-purity hexagonal boron nitride crystal. \textit{Phys. Rev. B} \textbf{94}, 121405(R) (2016).



\end{thebibliography}
\end{document}